\documentclass[fleqn,usenatbib]{mnras}

\usepackage{newtxtext,newtxmath}
\usepackage[T1]{fontenc}

\DeclareRobustCommand{\VAN}[3]{#2}
\let\VANthebibliography\thebibliography
\def\thebibliography{\DeclareRobustCommand{\VAN}[3]{##3}\VANthebibliography}

\usepackage{graphicx}	% Including figure files
\usepackage{amsmath}	% Advanced maths commands
\usepackage{xcolor}

\title[Low q MBHB Evolution in High Mach CBDs]{Unequal Mass Binary Evolution Driven by High Mach Circumbinary Disks}

\author[M. Clyburn and J. Zrake]{
Madeline Clyburn$^{1}$\thanks{E-mail: clyburn@clemson.edu} and
Jonathan Zrake$^{1}$\thanks{E-mail: jzrake@clemson.edu} \\
$^{1}$Department of Physics and Astronomy, Clemson University, 105 Sikes Hall, Clemson, SC 29634, USA\\
}

\date{Accepted XXX. Received YYY; in original form ZZZ}

\pubyear{\the\year{}}

\begin{document}
\label{firstpage}
\pagerange{\pageref{firstpage}--\pageref{lastpage}}
\maketitle

\begin{abstract}
We present a study of the gas-driven orbital evolution of unequal mass black hole binaries with circumbinary gas disks (CBDs), varying Mach number ($\mathcal{M}$) and viscosity ($\mathcal{\nu}$). Using two-dimensional grid-based hydrodynamics simulations spanning a thousand binary orbits at fixed separation, we explore low to moderate mass ratios ($q = 0.05$–$1.0$) and examine how variations in $\mathcal{M}$ and $q$ affect the torques and component accretion rates exerted by the CBD and consequently the binary evolution. Equal mass binary systems receive positive torques in low-$\mathcal{M}$ disks but transition to negative torques for $\mathcal{M} \gtrsim 25$. As $q$ decreases, the transition moves to higher Mach numbers. For $q<0.1$, we find no torque sign reversal below $\mathcal{M}\approx52$, except in sufficiently low-viscosity disks. We find that the secondary black hole cannot effectively repel the CBD, it instead accretes most of the inflowing gas from the CBD; these low mass ratio binaries in high viscosity disks therefore tend to outspiral, although inspiral can occur in less viscous environments. We also find that binaries with mass ratios in the range of $0.25 - 0.5$ can show preferential accretion favoring the primary when the gas viscosity is low, exemplifying an exception to the established rule of thumb that accretion favors the secondary. 
We discuss differences between our results and those reported in the literature on the orbital evolution and preferential accretion, and emphasize that our simulations extend into a regime that remains largely unexplored.
Overall, our results suggest that intermediate mass ratio inspirals (IMRIs) in CBDs may be less frequent, but this depends sensitively on the interplay between mass ratio, disk temperature, and viscosity.
\end{abstract}


%%%%%%%%%%%%%%%%% BODY OF PAPER %%%%%%%%%%%%%%%%%%%%%%%%%%%%%%%%%%%%%%%%%%%%%%%%%%%%%%%%%%%%%%%%%%%%

%%%%%%%%%%%%%%%%%%%%%%%%%%%%%%%%%%%%%%%%%%%%%%%%%%%%%%%%%%%%%%%%%%%%%%%%%%%%%%%%%%%%%%%%%%%%%%%%%%%%
% Intro
%%%%%%%%%%%%%%%%%%%%%%%%%%%%%%%%%%%%%%%%%%%%%%%%%%%%%%%%%%%%%%%%%%%%%%%%%%%%%%%%%%%%%%%%%%%%%%%%%%%%
\section{Introduction}
\label{sec:intro}

Most, if not all, massive galaxies host a central massive or supermassive black hole with mass $M \simeq 10^6-10^{10} \ $M$_{\odot}$ \citep[e.g.,][]{Dressler1988, Kormendy1995}. Additionally, major galaxy mergers occur frequently in our universe, and when two galaxies merge, their central black holes are expected to eventually form a massive black hole binary (MBHB) on timescales of $\sim$1 Gyr \citep{Begelman1980, Volonteri2003, Dotti2007, Khan2012, Kormendy2013}. These MBHBs will likely reside within the gas-rich environments of active galactic nuclei (AGN) \citep{Barnes1996} and may thus be surrounded by an accretion disk.

As the binary interacts with gas and stars most of these MBHBs are believed to contract until gravitational waves (GWs) drive the binary to coalescence \citep{Begelman1980}. The upcoming space-based GW detector, \textit{LISA}, will observe the GWs emitted by MBHBs in the final years to hours of their lives \cite[e.g.,][]{Sesana2021}. Through GW de-phasing, these GW detections may also reveal the presence of a circumbinary disk (CBD), offering a unique opportunity for multi-messenger astronomy  \citep[e.g.,][]{Hughes2002, Derdzinski2019,Derdzinski2021, IzquierdoVillalba2023,Tiede2024b}. Furthermore, identifying the electromagnetic (EM) counterparts to MBHB mergers will provide insight into the nature of accretion flows, the morphology of host galaxies, and the environments in which MBHBs evolve. A growing body of literature has begun to explore the expected EM signatures of MBHB inspirals for both equal and unequal mass systems \citep[see][]{Farris2010, Farris2011, Farris2012, Farris2015, Noble2012, Tang2018, Gutierrez2022, Avara2023, Krauth2023b, Ruiz2023, Cocchiararo2024, Franchini2024, Clyburn2025, Zrake2025}.

Nevertheless, despite decades of study, there remains significant uncertainty regarding whether binary–disk interactions can effectively shrink the binary orbit to separations where GWs can drive a merger within a Hubble time. This longstanding issue, commonly referred to as the ``final parsec problem'' \citep[e.g.,][]{Milosavljevic2003}, hinges on whether the CBD extracts or supplies angular momentum to the binary, thereby determining whether the system merges efficiently or stalls at sub-parsec scales.

Early analytical and numerical work generally supported the idea that gas torques remove angular momentum from the binary, accelerating inspiral and eventual coalescence \citep[e.g.,][]{Syer1995, Gould2000, Armitage2002, MacFadyen2008, Cuadra2009, Haiman2009, Shi2012, Kocsis2012, Roedig2012, Rafikov2016, Miranda2017, Franchini2021}. However, more recent simulations have challenged this view, showing that the binary net torque can become positive \citep[e.g.,][]{Munoz2019, Moody2019, Munoz2020, Duffell2020, Dittmann2021, DOrazio2021}. In this case, the binary gains angular momentum from the disk, leading to orbital expansion or even long-term stalling. 

\citet{Tiede2020} helped clarify this result by showing that the direction of angular momentum transfer depends sensitively on the disk’s structure. In particular, they found that equal mass ($q=1.0$) binaries in relatively warm disks ($\mathcal{M} = 10$) can experience positive torques and outspiral, while thinner and cooler disks ($\mathcal{M} \gtrsim 35$) tend to be negatively torqued and  thus, inspiral. Subsequent studies have reinforced this picture, demonstrating that the balance between inspiral and outspiral is controlled by a combination of parameters, including gas temperature, disk mass, viscosity, binary eccentricity, and mass ratio \citep[e.g.,][]{Heath2020, Franchini2021, Dittmann2022, Franchini2022, Penzlin2022, Siwek2023b, Dittmann2024, Tiede2025}. This sensitivity underscores the complex and still poorly understood role of disk–binary coupling in MBHB evolution. Understanding the physical conditions that govern the direction of angular momentum transfer is essential not only for determining whether gas-assisted mergers can solve the final parsec problem, but also for predicting the GW and EM signatures of MBHBs observable by \textit{LISA}.

To our knowledge, only two studies have begun to explore how binaries evolve in the low mass ratio, high Mach number regime. \citet{Penzlin2022} first investigated this parameter space, considering binaries with $q=0.1-1.0$ and disk Mach numbers up to $\mathcal{M} = 40$. Building on this, \citet{Dittmann2024} extended the study to mass ratios as low as $q = 0.01$, and disk Mach numbers up to $\mathcal{M} = 30$. Both works reported a similar torque trend to the equal mass case described in \citet{Tiede2020}: at sufficiently high Mach numbers, unequal mass binaries generally experience net negative torques and inspiral. However, in the mass ratio range of $0.01 \lesssim q \lesssim 0.05$, low mass ratio binaries were found to have net positive torques up to $\mathcal{M} = 30$ and thus outspiral \citep{Dittmann2024}.

Whether this outspiraling trend for low mass ratio binaries persists at higher Mach numbers ($\mathcal{M} > 30$) remains unknown. Yet this question is astrophysically significant, as observations indicate that AGN accretion disks are typically cooler and thinner, with $h/r \simeq 10^{-2} - 10^{-3}$, corresponding to Mach numbers in the range $\mathcal{M} \equiv (h/r)^{-1} \simeq 100 - 1,000$ \citep[e.g.,][]{Shakura1973, Krolik1999, Hubeny2001}. Moreover, a large fraction of the MBHB population is expected to have unequal masses \citep{Sesana2012, Bellovary2019}. Understanding how angular momentum is exchanged in this regime is crucial, as unequal mass binaries with thin, cool disks are predicted to dominate the observable MBHB population and future \textit{LISA} detections \citep[e.g., ][]{FrankKingRaine2002}.

In addition to binary migration behavior, the distribution of gas accretion between the two black holes could have substantial impact on the population of MBHBs in the universe. Simulations of binaries in relatively warm, low Mach number disks generally indicate that the secondary black hole accretes more efficiently than the primary, with preferential accretion factors $f \equiv \dot M_2 / \dot M_1 > 1$ \citep[e.g.,][]{Farris2014, Duffell2020, Munoz2020, Dittmann2021, Penzlin2022, Siwek2023b, Dittmann2024, Clyburn2025}. This trend is most pronounced for low mass ratio systems down to $q\approx 10^{-2}$, suggesting that accretion drives binaries toward equal masses over time. Consequently, many MBHBs within the \textit{LISA} detection horizon are expected to have near–equal mass ratios if this preferential accretion persists across all disk conditions.

However, the degree of preferential accretion appears sensitive to the viscosity prescription. \citet{Dittmann2024} found that binaries evolved with $\alpha$–viscosity CBDs exhibit higher $f$ values than those in disks with constant kinematic viscosity, implying that disk viscosity can alter which component dominates accretion. In our high Mach number simulations, we find that this trend reverses under certain conditions: binaries with moderate mass ratios ($q = 0.25$–$0.5$) embedded in either high Mach or low viscosity disks exhibit $f < 1$, indicating that the primary accretes more efficiently than the secondary (see Sec. \ref{subsec:Mdot-results}).

Our results for high viscosity cases show mild discrepancies with those of \citet{Dittmann2024}, who reported moderate secondary dominated accretion for $q = 0.5$ and $\mathcal{M} = 30$, while we find a modest preference for the primary in a similar configuration. This primary black hole accretion preference becomes more pronounced in low viscosity disks ($\nu = 10^{-4} \ \Omega_b a^2$), a parameter regime not yet systematically explored. Hints of such sensitivity are already evident in \citet{Dittmann2024}, who noted that the value of $f$ depends strongly on the adopted viscosity model. If accretion in nearly inviscid, high-Mach disks is biased toward the primary black hole, this bias may influence the long-term evolution of unequal-mass MBHBs. Instead of converging toward equal masses, binaries might maintain or even decrease in mass ratio over time. This outcome would suggest that unequal mass binaries may constitute a larger fraction of the \textit{LISA} detectable population than previously anticipated.

In this work, we expand upon the \citet{Dittmann2024} parameter space by probing significantly higher Mach numbers and similar low mass ratios with two-dimensional hydrodynamics simulations. Our results directly test the robustness of the inspiral–outspiral boundary and clarify how disk thickness and binary mass ratio jointly regulate angular momentum exchange between the binary and CBD. The remainder of this chapter is organized as follows. We begin in Sec. \ref{sec:theory} by outlining the physical framework governing MBHBs embedded in circumbinary accretion disks. Sec. \ref{sec:numerical} details the equations of motion, initial conditions, and numerical setup used to model these systems. Our main findings for the torque, accretion behavior, and orbital evolution are presented in Sec. \ref{sec:results}, and their broader implications for the MBHB population and \textit{LISA} detections are discussed in Sec. \ref{sec:summary}. Tests of the robustness of our results to numerical choices are provided in Appendix \ref{sec:appendix}.

%%%%%%%%%%%%%%%%%%%%%%%%%%%%%%%%%%%%%%%%%%%%%%%%%%%%%%%%%%%%%%%%%%%%%%%%%%%%%%%%%%%%%%%%%%%%%%%%%%%%
% Analytic section
%%%%%%%%%%%%%%%%%%%%%%%%%%%%%%%%%%%%%%%%%%%%%%%%%%%%%%%%%%%%%%%%%%%%%%%%%%%%%%%%%%%%%%%%%%%%%%%%%%%%
\section{Physical Picture}
\label{sec:theory}
Before presenting our simulation results, we first provide a brief theoretical overview of circumbinary accretion in MBHBs, focusing on the formation of accretion disks, the accretion dynamics, and the angular momentum transport within these systems.

\subsection{Circumbinary Accretion Disks}
\label{subsec:AGNdisk}
\begin{figure*}
  \includegraphics[]{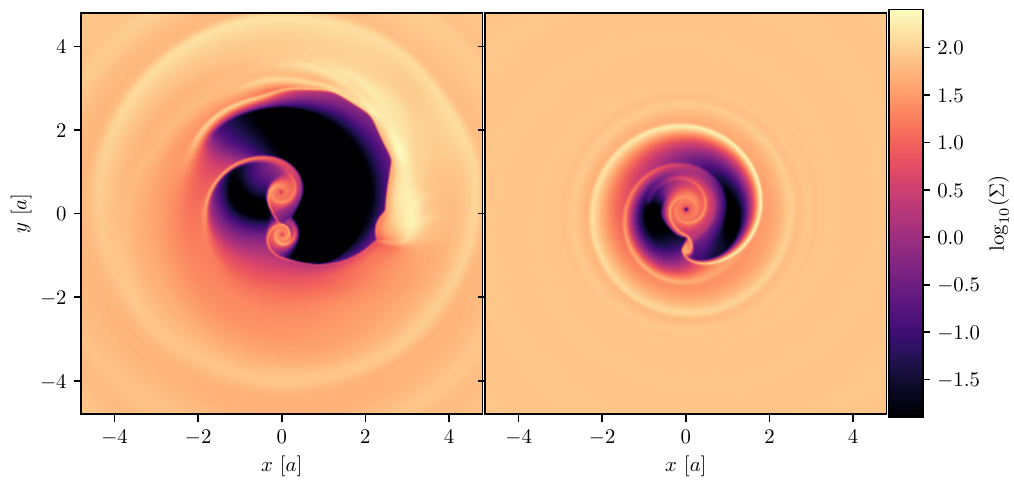}
  \caption{Plot of the two-dimensional surface density for a thin disk around an MBHB with mass ratio $q=1.0$ (left) and $q=0.1$ (right), accreting from a CBD. The image is from a simulation snapshot, using \texttt{Sailfish}, taken well after the disk is viscously relaxed. The axes show the $x-y$ plane in units of the binary separation $a$.}
  \label{fig:2dsig}
\end{figure*}
The physical picture of the relevant system is a black hole binary with total mass $M = M_1 + M_2$, semi-major axis $a$, and orbital frequency $\Omega_b = \sqrt{GM/a^3}$, embedded within an AGN disk. Far from the binary, mass flows through the disk at the rate $\dot M_{\infty}$ due to viscous transport. Gas flows inwards onto the binary at a rate $\dot M = \dot M_1 + \dot M_2$. Once the disk is viscously relaxed and a steady-state accretion flow is achieved, we can assume that the binary accretes at the inflow rate through the CBD: $\dot M = \dot M_{\infty}$. 

Gas flow near the binary depends sensitively on the system's mass ratio. In binaries with small mass ratios, the flow resembles that of a planet-forming disk. It consists of an outer CBD, a low-density annular gap around the secondary’s orbit, a prominent minidisk around the primary (more massive) BH, and a smaller minidisk around the secondary (less massive) BH. In contrast, equal mass binaries carve out a low-density cavity within the CBD, with equivalent minidisks around each BH. Another distinctive feature of equal mass systems is the formation of a dense, asymmetric “lump” at the cavity’s edge \citep[e.g.,][]{Roedig2011, Shi2012, Noble2012, Dorazio2013, Dorazio2016, Farris2014, Farris2015, Tang2018}. 

In Fig. \ref{fig:2dsig}, we show the two-dimensional surface density of a quasi-steady gas disk for both an equal mass and an unequal mass binary system, plotted out to a radius of $r = 4.5a$. The distinct morphologies of gas flow for binaries with different mass ratios can be seen in these simulation snapshots. The characteristic overdense “lump” near the cavity wall and the eccentric cavity in the CBD can be seen in the left-hand panel for a $q=1.0$ binary; the less eccentric and more symmetric gap in the CBD is found in the right-hand panel for a $q=0.1$ binary. These structural differences in the flow of gas around MBHBs can influence the exchange of angular momentum from the disk to the binary and the dynamics of accretion onto the black hole components.

\subsection{Angular Momentum Transport}
\label{subsec:torques}
For a circular MBHB as described in Sec. \ref{subsec:AGNdisk}, the orbital angular momentum is given by
\begin{equation}
J = \eta M \Omega_b a^2\ ,
\label{paper3-eqn:angmom}
\end{equation}
where $\eta \equiv q / (1+q)^2$, for binary mass ratio $q \equiv M_2 / M_1$. Differentiating Eqn. \ref{paper3-eqn:angmom} with respect to time yields a relationship between the torque on the binary and the rates of change of the mass ratio, total binary mass, and black hole separation:
\begin{equation}
\frac{\dot J}{J} = \frac{\dot \eta}{\eta} + \frac{3}{2}\frac{\dot M}{M} + \frac{1}{2}\frac{\dot a}{a} \, ,
\label{paper3-eqn:Jdot}
\end{equation}
where $\dot M$ is the accretion rate onto the binary and $\dot a = \dot a_{\rm gas}$ is the rate of change of the binary separation due to gas interactions. The term $\dot\eta/\eta$ can be expressed as
\begin{equation}
\frac{\dot \eta}{\eta} = \frac{1 - q}{1 + f} \bigg( \frac{f}{q} - 1\bigg) \frac{\dot M}{M} \, ,
\label{paper3-eqn:etadot}
\end{equation}
where $f \equiv \langle\dot M_2\rangle / \langle \dot M_1 \rangle$ is the preferential accretion parameter. A similar derivation for Eqn. \ref{paper3-eqn:Jdot} can be found in \cite{Munoz2019}, \cite{Dittmann2021}, and \cite{Penzlin2022}.

Typically, torques on the binary are normalized to the accretion rate such that,
\begin{equation}
\frac{\dot J}{\dot M} = \ell \Omega_b a^2 \ ,
\label{paper3-eqn:Jdot-Mdot}
\end{equation}
where $\ell$ is a dimensionless torque parameter determined by complex gas flows near the binary and measured by numerical simulations. The torque parameter is typically not greater in magnitude than about $10$, which is confirmed for binary mass ratio as low as $q\approx 10^{-2}$ \citep[e.g.,][]{Munoz2019, Munoz2020, Duffell2020, Duffell2024, Penzlin2022, Dittmann2022, Dittmann2024, Tiede2020, Clyburn2025, Tiede2025}.

The gas-induced torque $\dot{J}$ in Eqns. \ref{paper3-eqn:Jdot} and \ref{paper3-eqn:Jdot-Mdot} can be decomposed into two distinct angular momentum transport mechanisms
\begin{equation}
\dot J = \dot J_{\rm grav} + \dot J_{\rm acc} \ .
\label{paper3-eqn:torques}
\end{equation}
The gravitational torque $\dot J_{\rm grav}$ arises from tidal interactions between the binary and the surrounding CBD. The accretion torque $\dot J_{\rm acc}$ results from the angular momentum of the accreted gas being exchanged to the black holes. Accretion torques are always positive, adding angular momentum to the binary, and are written such that $\dot J_{\rm acc} \simeq \eta \dot M a^2 \Omega_b$. In contrast, the gravitational torques can be either positive or negative, depending on the disk properties. Specifically, when the orbital Mach number $\mathcal{M}$ is large the gravitational torque tends to become negative \citep[e.g.,][]{Tiede2020, Penzlin2022, Dittmann2024, Tiede2025}. 

\subsection{Binary Evolution}
\label{subsec:binary-evolve}

Using Eqns. \ref{paper3-eqn:Jdot}-\ref{paper3-eqn:Jdot-Mdot}, the rate of change of the semi-major axis due to angular momentum and mass transfer between the gas disk and binary is thus
\begin{equation}
\frac{\dot a}{a} = 2  \bigg[\frac{\ell}{\eta} - \frac{3}{2} - \frac{1-q}{1+f}\bigg(\frac{f}{q} -1\bigg) \bigg] \frac{\dot M}{M}\, ,
\label{paper3-eqn:adot}
\end{equation}
consistent with Eqn. 8 of \citet{Penzlin2022}. From Eqn. \ref{paper3-eqn:adot}, it is clear that the sign and magnitude of the torque parameter $\ell$ directly influences whether the binary inspirals or outspirals. If $\ell < 0$, then the binary will inspiral, but if $\ell > 0$ and sufficiently large enough to overcome ``$\dot M$ hardening'', then the binary will outspiral. Note that $\ell > 0$ does not in general imply $\dot{a} > 0$; this distinction is important for interpreting the plots that follow.

The critical $\ell$ value for inspiral, $\ell_{\rm crit}$, is found by setting Eqn. \ref{paper3-eqn:adot} to zero such that
\begin{equation}
 \ell_{\rm crit} = \eta  \bigg[\frac{3}{2} + \frac{1-q}{1+f}\bigg(\frac{f}{q} -1\bigg) \bigg] \, .
\label{paper3-eqn:ell-crit}
\end{equation}
When $\ell > \ell_{\rm crit}$, then the binary undergoes gas-induced outspiral. 
For equal mass binaries ($q=1$), the critical parameter is $\ell_{\rm crit} \simeq 3/8$ \citep{Tiede2020}. At lower mass ratio ($q=0.05$), this threshold increases to $\ell_{\rm crit} \simeq 0.67$, if $f\approx4.44$ as found in \citet{Dittmann2024}.

%%%%%%%%%%%%%%%%%%%%%%%%%%%%%%%%%%%%%%%%%%%%%%%%%%%%%%%%%%%%%%%%%%%%%%%%%%%%%%%%%%%%%%%%%%%%%%%%%%%%
% Numerical section,
%%%%%%%%%%%%%%%%%%%%%%%%%%%%%%%%%%%%%%%%%%%%%%%%%%%%%%%%%%%%%%%%%%%%%%%%%%%%%%%%%%%%%%%%%%%%%%%%%%%%
\section{Numerical Methods}
\label{sec:numerical}
The 2D hydrodynamics simulations presented in the following results were performed using the publicly available \texttt{Sailfish} code \citep{Zrake2024}. \texttt{Sailfish} is a GPU-accelerated, grid-based hydrodynamics code that employs a second-order Godunov solver. It is specifically designed to simulate binary–disk interactions and has been widely used in studies of binary accretion \citep[e.g.,][]{WesternacherSchneider2022, WesternacherSchneider2023, Krauth2023a, Krauth2023b, Duffell2024, Tiede2024a, DeLaurentiis2024, Zrake2025, Clyburn2025, Tiede2025}.

\subsection{Equations of Motion}
\label{subsec:Equations}
Our simulations are governed by the vertically averaged, time-dependent mass continuity and Navier–Stokes equations:
\begin{align}
    \label{paper3-eqn:NS1}
    &\frac{\partial \Sigma}{\partial t} + \mathbf{\nabla}\cdot(\Sigma\mathbf{v}) = \dot{\Sigma}_{\rm sink} \, , \\
    &\frac{\partial \Sigma \mathbf{v}}{\partial t} + \mathbf{\nabla } \cdot (\Sigma \mathbf{v}\mathbf{v} + P\,\mathbf{I} - \mathbf{T}_{\text{vis}}) = \dot{\Sigma}_{\rm sink} \mathbf{v} + \mathbf{F}_g \, .
    \label{paper3-eqn:NS2}
\end{align}
Here, $\Sigma$ is the vertically-integrated mass density of the gas disk, $\mathbf{v}$ is the gas velocity, $P \equiv \Sigma c_s^2$ is the vertically-integrated gas pressure, and $\mathbf{I}$ is the identity tensor (assuming isotropic pressure). For the simulations in this manuscript, we adopt a locally isothermal equation of state, where the sound speed is calculated according to
\begin{equation}
c_s^2 = -\phi / \mathcal{M}^2 \ ,
\label{paper3-eqn:cs}
\end{equation}
where $\mathcal{M} \equiv \varv_{\rm \phi}/c_s$ is the orbital Mach number of the disk and $\phi$ is the gravitational potential of the binary. Equivalently, the Mach number can be written in terms of the aspect ratio of the disk $h/r$ such that $\mathcal{M} \equiv (h/r)^{-1}$. Because we assume a locally isothermal equation of state, we do not solve an energy equation, and processes such as shock heating and radiative cooling are not included.

In Eqn. \ref{paper3-eqn:NS2}, $\mathbf{F}_g$ is the vertically integrated gravitational force density and is related to the softened gravitational potential by $\mathbf{F}_g = -\Sigma\,\nabla\phi$. The gravitational potential $\phi$ of the binary is then given by
\begin{equation}
\phi = -\frac{GM_1}{\sqrt{r_1^2 + r_s^2}} - \frac{GM_2}{\sqrt{r_2^2 + r_s^2}} \ ,
\label{paper3-eqn:potential}
\end{equation}
where $r_1$ and $r_2$ are the distances to each BH, and $r_s$ is the gravitational softening length, which ensures the potential remains finite at the positions of each component BH. We assume the total binary mass $M$ is much larger than the disk mass, allowing us to neglect disk self-gravity appropriately.

The ``sink'' term $\dot{\Sigma}_{\rm sink}$, which appears in Eqns. \ref{paper3-eqn:NS1} and \ref{paper3-eqn:NS2}, represents the local exchange of mass and momentum between the gas disk and binary components during accretion. It is given by:
\begin{align}
    \dot{\Sigma}_{\rm sink} = -\frac{\Sigma} {\tau_{\rm sink}} \bigg( e^{-r_1^2\, /\, 2r_{\rm sink}^2} + e^{-r_2^2\, /\, 2r_{\rm sink}^2} \bigg) \ .
    \label{paper3-eqn:sink}
\end{align}
In Eqn. (\ref{paper3-eqn:sink}), $r_{\rm sink}$ denotes the distance from a grid cell to each binary component, which we have written as $r_{\rm sink_i}$, with $i=1,2$ for the primary and secondary black hole respectively. The sink radius for each component is set to
\begin{equation}
r_{\rm sink_i} = 0.05 \ M_i \ \Omega_b a^2\ ,
\label{paper3-eqn:rsink}
\end{equation}
where $M_i$ is the mass of the ith black hole in code units. The sink rate $\tau_{\rm sink}^{-1} = 100 \ \Omega_b a^2$ controls the local flow of gas across the sink radius and is chosen to be short compared to the orbital timescale in order to efficiently remove gas within the sink region while minimizing perturbations to the surrounding flow. We have shown the sensitivity of our measurements to the sink paremeters in Appendix \ref{sec:appendix}. 

The components of the viscous stress tensor $\mathbf{T}_{\text{vis}}$ in Eqn. \ref{paper3-eqn:NS2} are given by $T^{ij}_{\text{vis}} = \nu \Sigma \left(\partial^j v^i + \partial^i v^j - \partial_k v^k \delta^{ij} \right)$. \texttt{Sailfish} supports both constant-$\nu$ and $\alpha$-viscosity prescriptions. For this study, we have adopted the constant-$\nu$ viscosity prescription with kinematic viscosity coefficient set to $\nu = \sqrt{2} \times 10^{-3} \ \Omega_b a^2$, consistent with \cite{Tiede2020} and $\nu = 10^{-4} \ \Omega_b a^2$. We find that the torque measurements are significantly dependent on the choice of viscosity prescription and magnitude. We have shown the sensitivity of our measurements to the viscosity in Sec. \ref{sec:results} and Appendix \ref{sec:appendix}.

\subsection{Simulation Setup and Diagnostics}
\label{subsec:initial}
All simulations are initialized with a circular binary ($e = 0$) of fixed separation $a$ and orbital frequency $\Omega_b$ embedded in an approximate steady-state gas disk. The disk is initially circular and extends to a finite outer radius of $r = 32a$; see Fig. \ref{fig:2dsig} for 2D surface densities of simulation setups. The azimuthal velocity of the gas is initialized to the Keplerian profile, $v_{\phi} = \sqrt{GM (1-\mathcal{M}^{-2}) / r}$, and the radial velocity is set to the viscous drift speed, $v_r = -3\nu / 2r$. The corresponding initial surface density profile is given by
\begin{equation}
\Sigma_0(r) = \frac{\dot M_0}{3 \pi \nu(r)} \, ,
\label{paper3-eqn:steady-state-sigma-zero-torque}
\end{equation}
where $\dot M_0=\dot M_{\infty}$, the inflow rate at $r=\infty$. This expression corresponds to a steady-state disk with zero net angular momentum flux. Following initialization, the disk requires several viscous timescales at $r \approx a$ to relax to a quasi-steady state. For reference, when $\nu = \sqrt{2} \times 10^{-3} \ \Omega_b a^2$, the viscous timescale at $r=a$ is $t_{\rm visc}(a) \equiv 2 a^2/ 3 \nu \simeq 75 \ {\rm orbits}$. As such, we exclude the first $\sim100$ orbits from our analysis to avoid contamination from these initial transients. Each simulation runs for a total duration of 1000 binary orbits which we have found is long enough to reach a quasi-steady state (see Sec. \ref{subsec:Mdot-results} and Appendix \ref{sec:appendix}).

The accretion rate onto each black hole is measured by separately integrating the two terms in Eqn. \ref{paper3-eqn:sink} corresponding to the primary and secondary black hole accretion rates. The time series of the gravitational torque is computed by integrating the gravitational torque density over the entire surface of the disk,
\begin{equation}
    \dot J_{\rm grav} = \int r \times \mathbf{F}_g \ dA \, .
    \label{eq:Jdot_grav}
\end{equation}
The gravitational torque can also be decomposed into the contributions from the inner ($r<a$) and outer ($r>a$) disks. The time series of the accretion torque is computed by integrating the sink term in Eqn. \ref{paper3-eqn:sink} multiplied by the gas speed $\mathbf{v}$,
\begin{equation}
    \dot J_{\rm acc} = \int \mathbf{r} \times (\dot{\Sigma}_{\rm sink} \mathbf{v}) \ dA \cdot \hat{z}\, .
    \label{eq:Jdot_acc}
\end{equation}

The domain size of our Cartesian mesh simulations is $r_{\rm out} = 32 a$, and we use a grid spacing of $\Delta x = 0.01 a$ for a majority of the results below. With such grid spacing and domain size we consider $(2r_{\rm out} / \Delta x c)^2 = 6400\times6400$ grid zones. In Appendix \ref{sec:appendix} we test the sensitivity of our torque measurements to the domain size and grid spacing and find that $r_{\rm out} = 32 a$ and $\Delta x = 0.01 a$ are sufficient for numerical convergence of our torque measurements.

%%%%%%%%%%%%%%%%%%%%%%%%%%%%%%%%%%%%%%%%%%%%%%%%%%%%%%%%%%%%%%%%%%%%%%%%%%%%%%%%%%%%%%%%%%%%%%%%%%%%
% Results section
%%%%%%%%%%%%%%%%%%%%%%%%%%%%%%%%%%%%%%%%%%%%%%%%%%%%%%%%%%%%%%%%%%%%%%%%%%%%%%%%%%%%%%%%%%%%%%%%%%%%
\section{Simulation Results}
\label{sec:results}
We perform hydrodynamics simulations of black hole binaries with mass ratios ranging from $q = 0.05 - 1.0$, embedded in CBDs with Mach numbers $\mathcal{M} = 10 - 52$, evolved for $1,000$ binary orbits. Below, we outline our simulation results across this parameter space, focusing on the high Mach number and low mass ratio regimes. 

\subsection{Torques}
\label{subsec:Jdot-results}
\begin{figure}
\centering
  \includegraphics[]{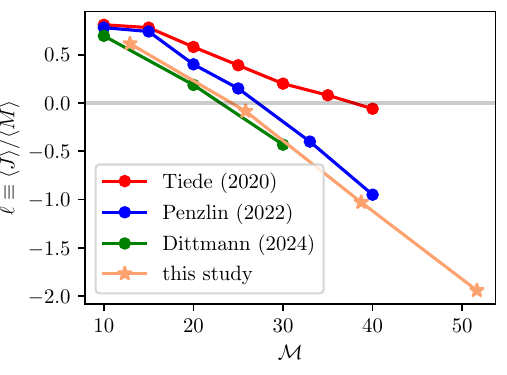}
  \caption{Plot of the dimensionless torque parameter, $\ell$, as a function of Mach number, $\mathcal{M}$, for an MBHB with mass ratio $q = 1.0$. Results are compared with those from \protect\cite{Tiede2020}, \protect\cite{Penzlin2022} and \protect\cite{Dittmann2024}. At higher Mach numbers, the torques become increasingly negative, in agreement with the trends reported in these previous studies.}
  \label{fig:ell-compare}
\end{figure}
Fig. \ref{fig:ell-compare} shows the average dimensionless torque parameter, $\ell \equiv \langle \dot J \rangle / \langle \dot M \rangle$, as a function of Mach number for an equal mass MBHB\footnote{$\langle \rangle$ denotes a time-average over multiple viscous timescales.}. Our results are compared with those of \cite{Tiede2020}, \cite{Penzlin2022}, and \cite{Dittmann2024}. We find very close agreement with \cite{Dittmann2024} and good agreement with \cite{Penzlin2022} at high Mach numbers. The discrepancies with \cite{Tiede2020} may result from their slightly lower resolution, as discussed further in \cite{Penzlin2022}. 

\subsubsection{High-$\nu$ Torques}
\label{subsubsec:high-nu-torque}
\begin{figure}
\centering
  \includegraphics[]{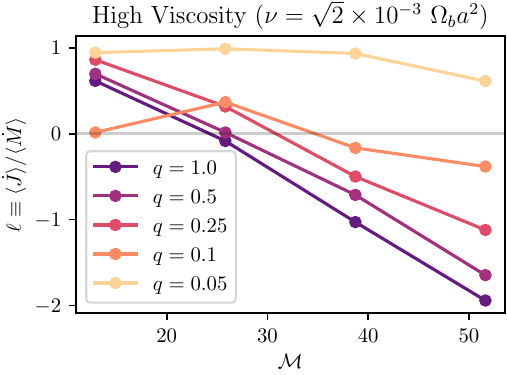}
  \caption{Plot of the dimensionless torque parameter, $\ell$, as a function of Mach number, $\mathcal{M}$, for MBHBs with mass ratio in the range $q = 0.05-1.0$. Results shown here are for the highest resolution tested ($\Delta x = 0.005 a$).}
  \label{fig:ell-q}
\end{figure}
\begin{figure*}
  \includegraphics[]{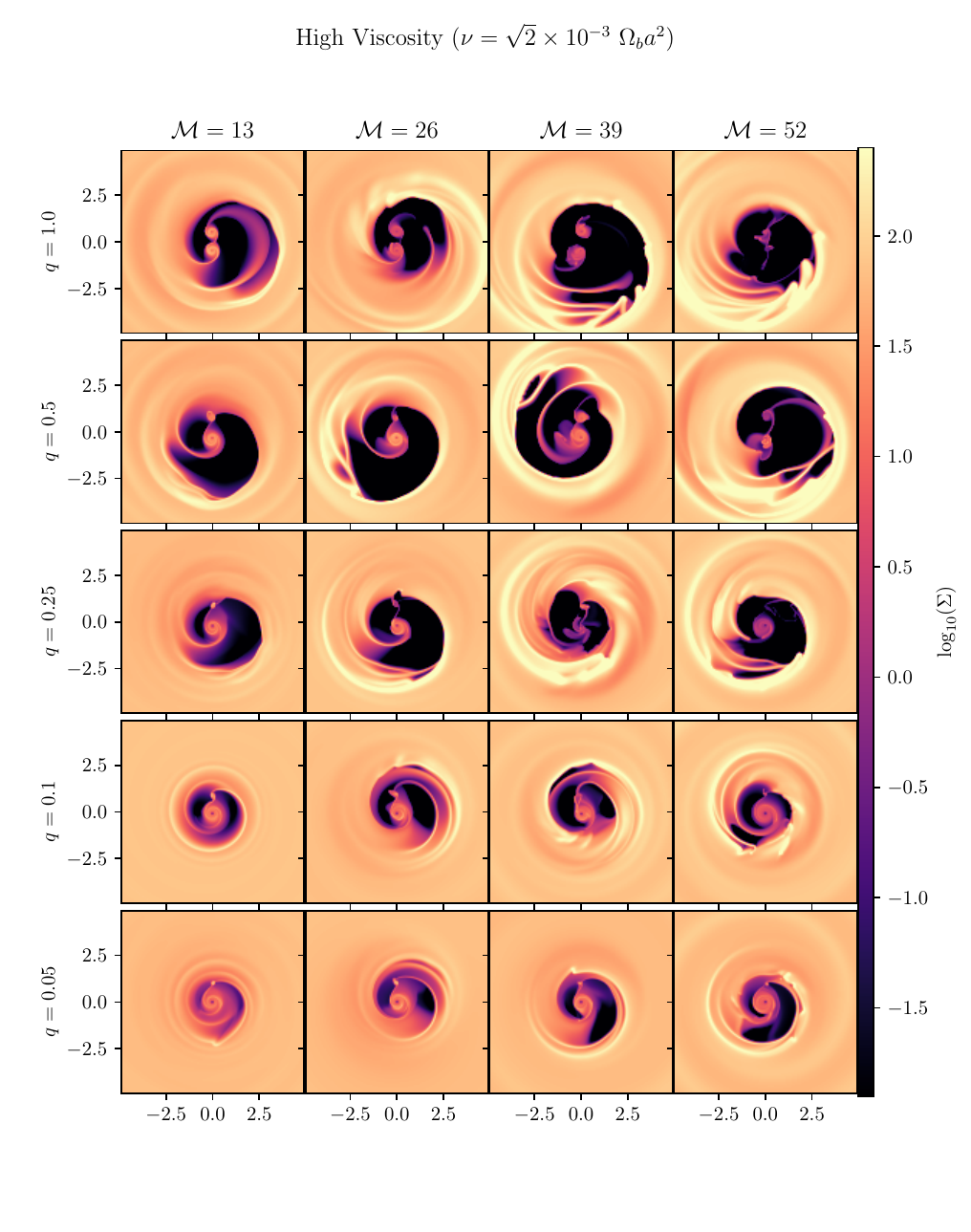}
  \caption{Plot of the two-dimensional surface density for MBHBs with varying Mach numbers and mass ratios. From top to bottom, rows show decreasing mass ratio in the range $q=0.05-1.0$. From left to right, columns show increasing Mach number in the range $\mathcal{M} = 13-52$.}
  \label{fig:sig-q}
\end{figure*}
%
%\begin{figure}
%\centering
%  \includegraphics[]{Figures/JdotMachMassRatio.pdf}
%  \caption{Plot of the average torques as a function of Mach number, $\mathcal{M}$, for MBHBs with mass ratios in the range $q=0.05-1.0$. The color for each line corresponds to the mass ratio of the binary, same as in Fig. \ref{fig:ell-q}. The gravitational torque from the CBD ($r>a$) is plotted in the top panel, the gravitational torque from the minidisk region ($r<a$) is plotted in the middle panel, and the accretion torque is plotted in the bottom panel.}
% \label{fig:jdot}
%\end{figure}
%
All measurements of $\ell$ for equal mass binaries show that increasing the Mach number of the gas causes the average torque parameter to decrease and eventually become negative. To test whether this trend persists in unequal mass binaries, we plot $\ell$ for binaries spanning a range of Mach numbers and mass ratios in Fig. \ref{fig:ell-q}. For high mass ratio systems, the torque parameter follows a similar trend to that of equal mass binaries, transitioning to negative values at sufficiently high Mach numbers. As the mass ratio decreases, this positive-to-negative torque transition shifts to higher Mach numbers. However, for $q \lesssim 0.1$, this transition is not seen even up to $\mathcal{M} \simeq 52$ and the disk fails to produce net negative torques. As seen in Fig. \ref{fig:ell-q}, for $q=0.05$ the torque parameter remains positive up to the highest Mach number tested. Although $\ell$ shows a slight downward trend at $\mathcal{M} \simeq 52$, this suggests that there could be a transition to negative binary torque at $\mathcal{M} > 52$.

Fig. \ref{fig:sig-q} presents the two-dimensional surface density distributions for binaries with varying mass ratios and disk Mach numbers (matching the parameter space in Fig. \ref{fig:ell-q}). At high mass ratios, the circumbinary cavity is larger, more eccentric, and more depleted of gas compared to systems with low mass ratios. Due to the negative binary torque in high Mach number disks, gas is flung outward from the cavity while viscous inflow through the CBD continues. This produces a gas pile-up at the inner edge of the CBD, more pronounced in near equal mass systems. While a similar feature appears for unequal mass binaries, it becomes less pronounced as the mass ratio decreases.

%Fig. \ref{fig:jdot} shows the time-averaged torques as a function of Mach number, decomposed into three components: the gravitational torque from the CBD ($r > a$, top panel), the gravitational torque from the minidisk and cavity region ($r < a$, middle panel), and the accretion torque (bottom panel). The CBD gravitational torques are consistently negative and vary little with mass ratio. In contrast, the minidisk gravitational torque is positive and shows a pronounced enhancement for $q = 0.05$, where its magnitude is approximately 2–4 times larger than in higher mass ratio systems. Additionally, accretion torques remain strong at high Mach numbers for low-$q$ binaries. Together, the amplified minidisk and accretion torques likely offset the negative CBD contributions, preventing the $q=0.05$ binary from experiencing a net negative torque.

%An apparent outlier in Fig. \ref{fig:jdot} corresponds to the equal mass binary with a disk Mach number of $26$. Here, the accretion torque and the excised gravitational torque are significantly lower than in comparable systems, while the minidisk gravitational torque is anomalously high. Additional simulations at higher resolution will be required to assess whether this reflects a genuine physical effect or a numerical artifact, but is beyond the scope of this paper.

\subsubsection{Low-$\nu$ Torques}
\label{subsubsec:low-nu-torque}
\begin{figure}
\centering
  \includegraphics[]{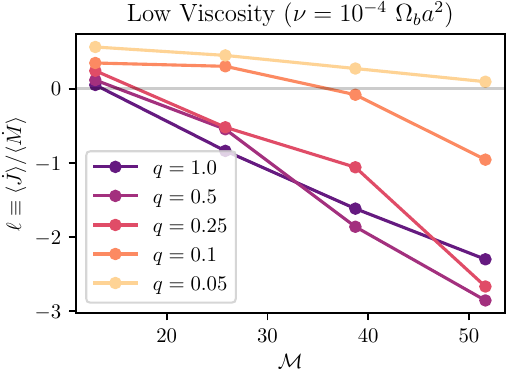}
  \caption{Plot of the dimensionless torque parameter, $\ell$, as a function of Mach number, $\mathcal{M}$, for an MBHB with varying mass ratio. The plot is similar to Fig. \ref{fig:ell-q} but with lower viscosity. The color for each line corresponds to the mass ratio of the binary, same as in Fig. \ref{fig:ell-q}.}
  \label{fig:ell-visc-q}
\end{figure}
\begin{figure*}
  \includegraphics[]{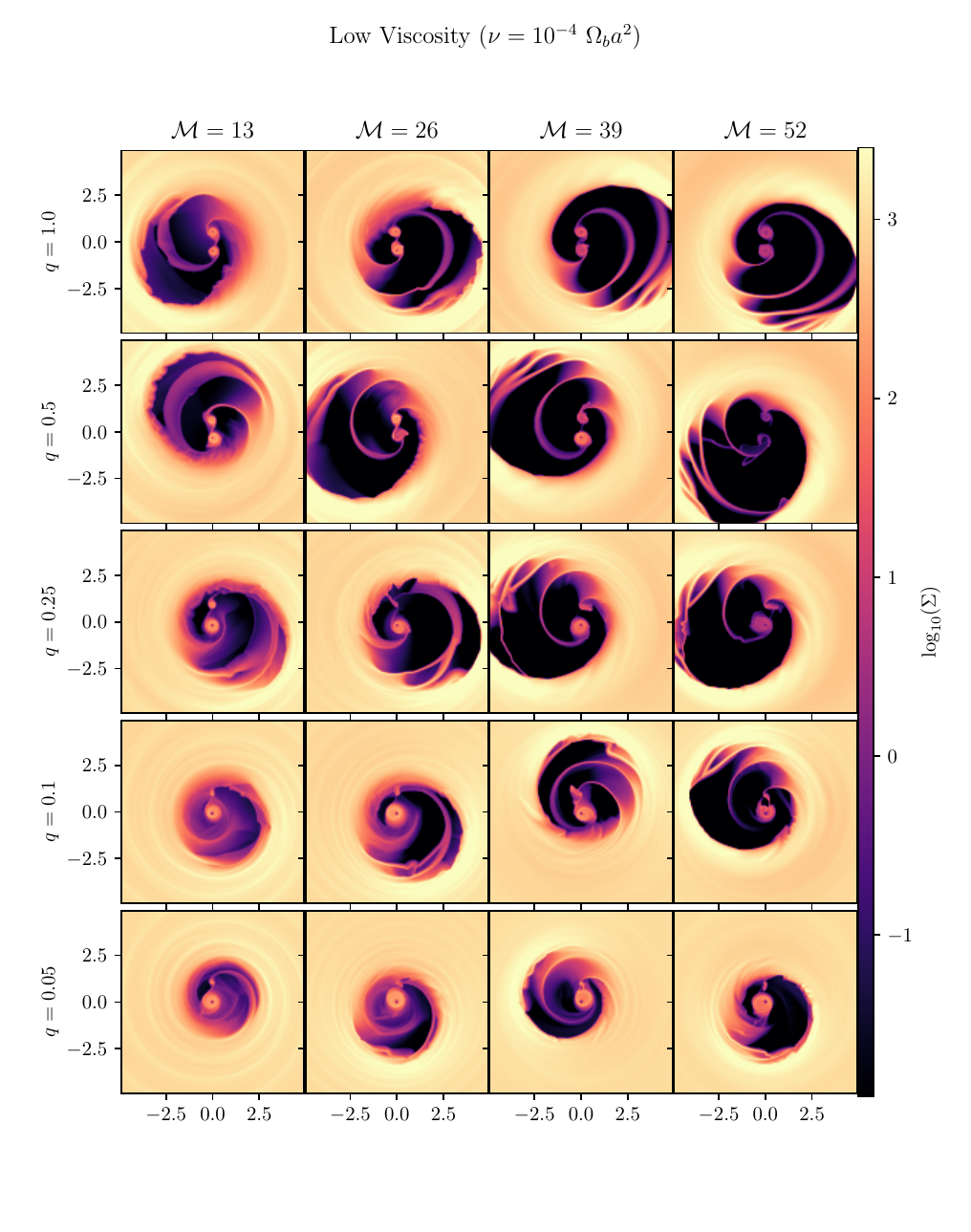}
  \caption{Plot of the two-dimensional surface density for MBHBs with varying Mach numbers and mass ratios, similar to Fig. \ref{fig:sig-q} but with lower viscosity. From top to bottom, rows show decreasing mass ratio in the range $q=0.05-1.0$. From left to right, columns show increasing Mach number in the range $\mathcal{M} = 13-52$.}
  \label{fig:sig-visc}
\end{figure*}
%
%\begin{figure}
%\centering
%  \includegraphics[]{Figures/JdotMachViscNu.pdf}
%  \caption{Plot of the average torques as a function of Mach number, $\mathcal{M}$, for MBHBs with varying mass ratio. The plot is similar to Fig. \ref{fig:jdot} but with lower viscosity. The color for each line corresponds to the mass ratio of the binary, same as in Fig. \ref{fig:ell-q}. The gravitational torque from the CBD ($r>a$) is plotted in the top panel, the gravitational torque from the minidisk region ($r<a$) is plotted in the middle panel, and the accretion torque is plotted in the bottom panel.}
%  \label{fig:jdot-visc-nu}
%\end{figure}
%
We now test the sensitivity of our results to the normalization of the kinematic viscosity. Fig. \ref{fig:ell-visc-q} shows the dependence of $\ell$ on $q$ and $\mathcal{M}$ for a lower-viscosity CBD. In the section above (Sec. \ref{subsubsec:high-nu-torque}), we chose a relatively high viscosity $\nu = \sqrt{2} \times 10^{-3} \ \Omega_b a^2$, consistent with \citet{Tiede2020} and \citet{Penzlin2022}. Reducing the viscosity to $\nu = 10^{-4} \ \Omega_b a^2$ causes the average torque parameter to become systematically more negative, decreasing by a factor $\sim2.5$ for binaries with $q=0.25 - 1.0$. At $q=0.1$, $\ell$ shows little sensitivity to viscosity, while for $q=0.05$ the torque decreases by $\sim 50\%$. This suggests that low mass ratio binaries could be driven to inspiral in sufficiently low viscosity disks. 

In Fig. \ref{fig:sig-visc}, we show the two-dimensional surface density for binaries embedded in less viscous disks. As in Fig. \ref{fig:sig-q}, the cavity becomes increasingly depleted at higher Mach numbers, while streams of gas are expelled from the binary toward the outer disk, enhancing the pileup at the cavity’s inner edge. In contrast, the surface density near the inner edge of the CBD is roughly an order of magnitude higher in low viscosity disks as compared to the highly viscous cases shown in Fig. \ref{fig:sig-q}.

%Fig. \ref{fig:jdot-visc-nu} shows the time-averaged torques as a function of Mach number, decomposed as in Fig. \ref{fig:jdot}, but for the lower-viscosity case. Comparing these two figures, the average CBD torque becomes more negative while the average minidisk torque increases at lower viscosity. This trend aligns with prior results showing that negative gravitational torques strengthen as the magnitude of $\nu$ decreases \citep[e.g.,][]{Tiede2020,Penzlin2022,Dittmann2022}.

\subsection{Accretion Rates}
\label{subsec:Mdot-results}
\begin{figure}
\centering
  \includegraphics[]{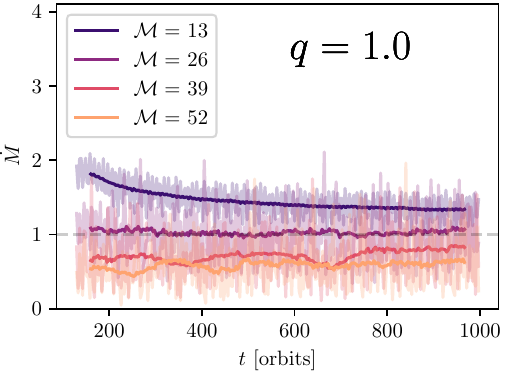}
  \caption{Timeseries of the total binary accretion rate $\dot M$ onto an equal mass binary for disks with varying Mach numbers. Transparent curves highlight the stochastic variability of the accretion rates.}
  \label{fig:mdot-1q}
\end{figure}
\begin{figure}
\centering
  \includegraphics[]{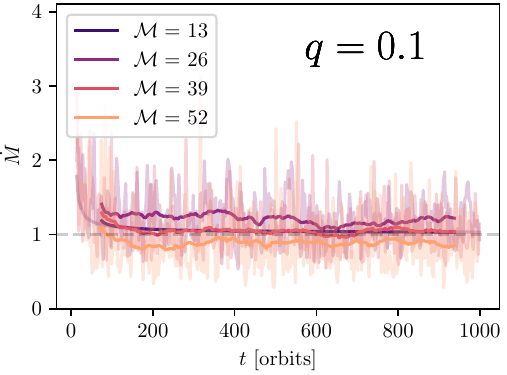}
  \caption{Timeseries of the total binary accretion rate $\dot M$ onto an unequal mass binary ($q=0.1$) for disks with varying Mach numbers. Transparent curves highlight the stochastic variability of the accretion rates.}
  \label{fig:mdot-01q}
\end{figure}

Fig. \ref{fig:mdot-1q} shows the total accretion rate onto an equal mass binary for disks with varying Mach numbers. The system reaches a steady state within the first $\sim 200$ orbits. Enhanced accretion at low Mach numbers and suppressed accretion at high Mach numbers arise from the torques exerted by the disk on the binary: positive torques drive an initial enhancement in accretion, while negative torques cause an initial suppression \citep{Zrake2025}. This suppression for high Mach number, equal mass binaries is consistent with the findings of \cite{Tiede2025}.

Fig. \ref{fig:mdot-01q} shows the total accretion rate onto an unequal mass binary ($q=0.1$) for disks with varying Mach numbers. As in the equal mass case, the system reaches a steady state within the first few hundred orbits. However, unlike $q=1.0$ systems, the accretion rates show little sensitivity to Mach number. Notably, for $q=0.1$, negative torques are observed only for disks with $\mathcal{M} \gtrsim 40$. At even lower mass ratios, this trend becomes more pronounced, with the disk exerting exclusively positive torques up to at least $\mathcal{M} \simeq 52$.

\subsubsection{High-$\nu$ Accretion Rates}
\label{subsubsec:high-nu-accretion}
\begin{figure}
\centering
  \includegraphics[]{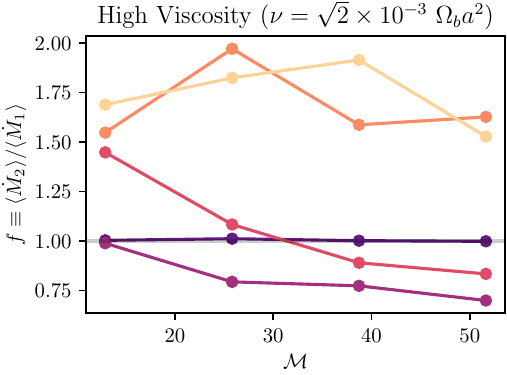}
  \caption{Plot of the average preferential accretion rate $f \equiv \langle\dot M_2\rangle /\langle\dot M_1\rangle$ as a function of Mach number, $\mathcal{M}$, for MBHBs with mass ratios in the range $q=0.05-1.0$. The color for each line corresponds to the mass ratio of the binary, same as in Fig. \ref{fig:ell-q}.}
  \label{fig:pref-acc}
\end{figure}

Fig. \ref{fig:pref-acc} shows the average preferential accretion rate, $f \equiv \langle\dot{M}_2\rangle / \langle\dot{M}_1\rangle$, for binary-disk systems spanning a range of Mach numbers and mass ratios. As in previous studies \citep[e.g.,][]{Farris2014,Duffell2020,Munoz2020,Dittmann2021,Siwek2023b,Clyburn2025}, preferential accretion onto the secondary black hole increases as the mass ratio decreases. Our results differ slightly from those in \cite{Duffell2020}, \cite{Munoz2020}, and \cite{Siwek2023a}, who find $f \approx 8$ for $q=0.1$. Instead, our findings are more consistent with \cite{Farris2014} and \cite{Dittmann2021}, where $f \approx 2$ for $q=0.1$. A possible explanation for these differences is our treatment of viscosity. Both \cite{Farris2014}, \cite{Dittmann2021} and our results here adopt a constant viscosity prescription, while the studies with vastly different results for the preferential accretion rate use $\alpha$-viscosity \citep{Siwek2023b}. 

At higher Mach numbers in Fig. \ref{fig:pref-acc}, binaries with $q = 0.25$ and $q = 0.5$ exhibit the opposite trend seen at lower mass ratios, where the primary black hole accretes more than the secondary. This behavior is also visible in Fig. \ref{fig:sig-q}, where the secondary’s minidisk is more depleted than the primary’s for $q = 0.25$–$0.5$ at high Mach. Our finding contrasts with the results of \citet{Dittmann2024}, who report secondary-dominated accretion for similar mass ratios at Mach 30. The discrepancy may arise from differences in the numerical implementation. Our simulations employ standard sink prescriptions, whereas theirs use torque-free sinks, which can alter the preferential accretion rate. Additionally, \citet{Dittmann2024} found moderate dependence of their results on sink rate, which could explain our results' discrepancies as well.

\subsubsection{Low-$\nu$ Accretion Rates}
\label{subsubsec:low-nu-accretion}
\begin{figure}
\centering
  \includegraphics[]{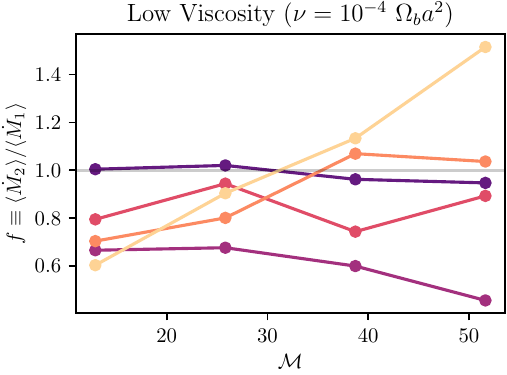}
  \caption{Plot of the average preferential accretion rate $f \equiv \langle\dot M_2\rangle /\langle\dot M_1\rangle$ as a function of Mach number, $\mathcal{M}$, for MBHBs with varying mass ratios. The plot is similar to Fig. \ref{fig:pref-acc} but with lower viscosity. The color for each line corresponds to the mass ratio of the binary, same as in Fig. \ref{fig:ell-q}.}
  \label{fig:f-visc-nu}
\end{figure}

Fig. \ref{fig:f-visc-nu} presents the dependence of the average preferential accretion rate, $f$, on Mach number and mass ratio for the low viscosity configuration described in Sec. \ref{subsubsec:low-nu-torque}. With lower viscosity, the observed accretion trends reverse relative to those in Fig. \ref{fig:pref-acc}. For binaries with near-equal masses ($q \geq 0.25$), accretion favors the primary black hole across all Mach numbers examined. In contrast, for binaries with low mass ratios ($q \leq 0.1$), the primary black hole dominates accretion only at low Mach numbers, while the system transitions to secondary-dominated accretion as the Mach number increases.

We propose the following explanation for the observed accretion behavior in low viscosity CBDs. The secondary minidisk, which orbits the less massive BH, is replenished quasi-periodically by gas streams falling inward from the CBD. When the secondary is located on the far side of the cavity, its disk viscously drains onto the secondary black hole, reducing the disk mass at a rate set by the local viscous rate. During subsequent inflow events from the CBD, part of the incoming gas overshoots or “skips” off the outer edge of the secondary disk, falling instead onto the primary minidisk. If the secondary disk has drained substantially between feeding episodes, it can accommodate more of the inflowing gas. However, when the viscosity is low, the secondary disk drains less efficiently, remains more massive, and therefore rejects a larger fraction of the incoming material, thus, potentially enhancing accretion onto the primary black hole.

\subsection{Binary Evolution}
\label{subsec:evolution-results}

The long-term evolution of MBHBs interacting with a CBD is described by Eqn. \ref{paper3-eqn:adot}, where the semi-major axis evolution depends on both torque and relative accretion rates. Secs. \ref{subsec:Jdot-results} and \ref{subsec:Mdot-results} show how these quantities vary with mass ratio, gas temperature (Mach number), and viscosity. Most notably, Fig. \ref{fig:ell-q} demonstrates that low mass ratio binaries ($q = 0.05$) experience net positive torques at high viscosity even up to $\mathcal{M} \simeq 52$, in contrast to near equal mass binaries and systems in less viscous disks. This indicates that low-$q$ systems may have different evolutionary fates depending on the viscosity of their surrounding CBD. 

\subsubsection{High-$\nu$ Binary Evolution}
\label{subsubsec:high-nu-adot}
\begin{figure}
\centering
  \includegraphics[]{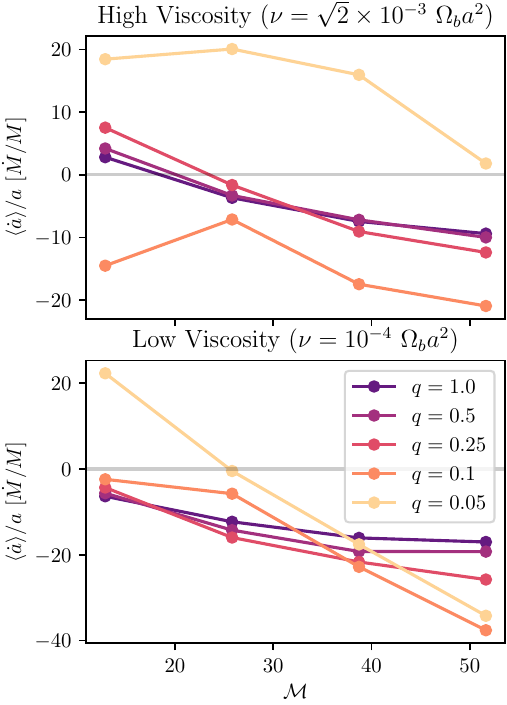}
  \caption{Plot of the average binary separation rate of change, $\langle \dot a \rangle / a$, due to interactions with the gas disks as a function of Mach number, $\mathcal{M}$, for MBHBs with mass ratios in the range $q=0.05-1.0$. The color for each line corresponds to the mass ratio of the binary, same as in Fig. \ref{fig:ell-q}. The top panel is for simulations with high-$\nu$ and the bottom panel is for the simulations with low-$\nu$}
  \label{fig:adot}
\end{figure}

Fig. \ref{fig:adot} shows the average semi-major axis evolution $\langle \dot a \rangle$, computed from Eqn. \ref{paper3-eqn:adot}, for the same range of $q$ and $\mathcal{M}$. In agreement with previous studies, near equal mass binaries outspiral for $\mathcal{M} \lesssim 20$ but inspiral at higher Mach numbers \citep[e.g.,][]{Tiede2020,Penzlin2022,Dittmann2024}. For $q = 0.1$, the binary inspirals across all $\mathcal{M}$ tested, consistent with results at $\mathcal{M} = 10$ in the current literature \citep[e.g.,][]{Siwek2023b}. At still lower mass ratios, the trend reverses, and the binary exhibits positive $\langle\dot{a}\rangle$, corresponding to outspiral. For $q = 0.05$, the semi-major-axis evolution approaches zero, suggesting that sufficiently cold disks could eventually drive inspiral. However, these results imply that binaries with $q \lesssim 0.05$ may never reach separations that induce GW inspiral, or that such inspiral may be significantly delayed.

\subsubsection{Low-$\nu$ Binary Evolution}
\label{subsubsec:low-nu-adot}

Fig. \ref{fig:adot} shows the dependence on $\mathcal{M}$ of the secular migration rate, $\langle \dot{a} \rangle / a$, for the lower viscosity setup described in Secs. \ref{subsubsec:low-nu-torque} and \ref{subsubsec:low-nu-accretion}. We find that, regardless of Mach number, higher mass ratio binaries ($q \ge 0.1$) undergo inspiral when the viscosity is sufficiently low ($\sim10^{-4}$), and that $q = 0.05$ binaries also inspiral for $\mathcal{M} \gtrsim 25$. This behavior mirrors the results of \citet{Dittmann2022} (see their Fig. 3), where lower viscosity led to progressively faster inspiral for equal mass binaries. Here we show that this trend extends to unequal-mass systems as well. Lower viscosity promotes inspiral primarily by reducing the efficiency of angular momentum transport via viscous stresses. With smaller $\nu$, the CBD cannot replenish the cavity as quickly, so the central cavity remains depleted and the positive torques from accretion streams are weakened \citep{Dittmann2022}. The near-constancy of the $q = 0.1$ results with varying $\nu$, contrasted with the strong $\nu$-dependence at both higher and lower mass ratios, suggests that $q \simeq 0.1$ marks a transitional regime, but is beyond the scope of this work and additional simulations at higher resolution are needed to test this hypothesis.

%%%%%%%%%%%%%%%%%%%%%%%%%%%%%%%%%%%%%%%%%%%%%%%%%%%%%%%%%%%%%%%%%%%%%%%%%%%%%%%%%%%%%%%%%%%%%%%%%%%%
% Summary section
%%%%%%%%%%%%%%%%%%%%%%%%%%%%%%%%%%%%%%%%%%%%%%%%%%%%%%%%%%%%%%%%%%%%%%%%%%%%%%%%%%%%%%%%%%%%%%%%%%%%
\section{Summary}
\label{sec:summary}
\subsection{Main Results}
\label{subsec:main-results}
\begin{figure*}
\centering
  \includegraphics[]{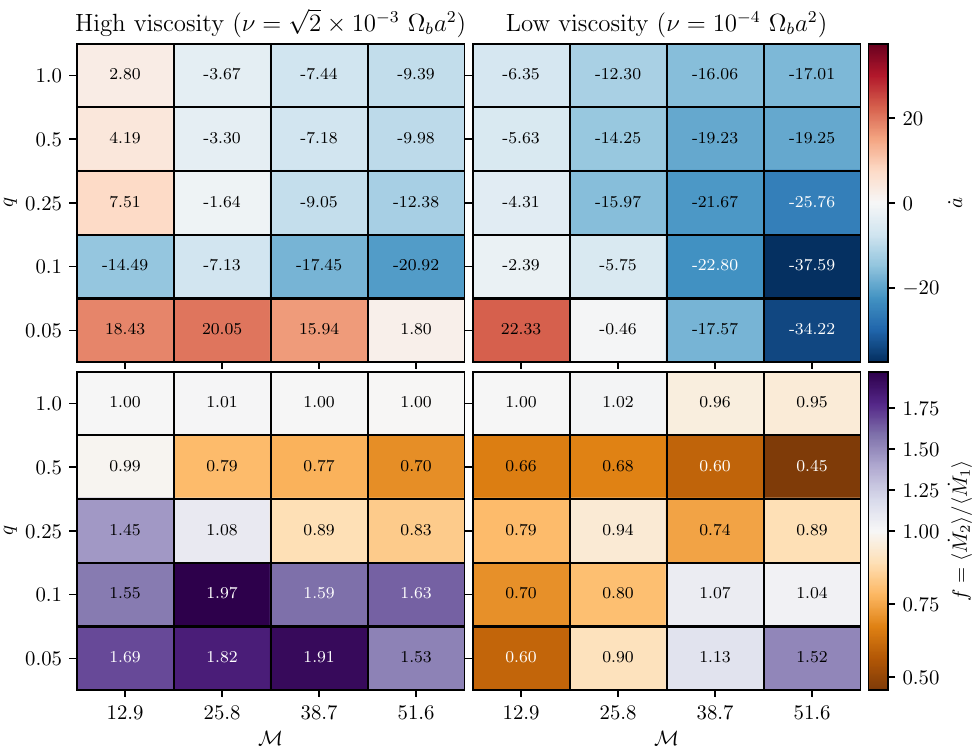}
  \caption{ 
Summary of binary migration and preferential accretion across parameter space. 
Top panels show the orbit-averaged migration rate $\dot{a}$ as a function of mass ratio $q$ and disk Mach number $\mathcal{M}$ for two values of the viscosity. 
Bottom panels show the preferential accretion parameter $f$. 
For $\dot{a}$, red indicates outspiral ($\dot{a}>0$), blue indicates inspiral ($\dot{a}<0$), and white corresponds to $\dot{a}\approx0$. 
For $f$, purple indicates preferential accretion onto the primary, orange indicates preferential accretion onto the secondary, and white corresponds to approximately equal accretion. 
Each cell shows the measured value from the simulations.
}
  \label{fig:summary}
\end{figure*}

We have performed numerical hydrodynamics simulations of unequal mass MBHBs, $q = 0.05 - 1.0$, embedded in high Mach number disks, $\mathcal{M} = 10 - 52$. The aim of this work is to better understand how thinner, cooler CBDs influence the orbital evolution and accretion dynamics of unequal mass MBHB inspirals. The main results of our study can be found in Fig. \ref{fig:summary} and are summarized below:
\begin{enumerate}
  \item For low mass ratio binaries ($q = 0.05$), we find that net positive torques persist across the entire Mach number range investigated. This sustained outspiral is primarily driven by gravitational torques from the minidisks within $r < a$, as the secondary black hole preferentially accretes most of the inflowing gas but fails to repel the CBD. As shown in Fig. \ref{fig:ell-q}, the torque remains positive up to $\mathcal{M} \simeq 52$ for high viscosity disks. However, in less viscous disks this trend breaks down, with Fig. \ref{fig:ell-visc-q} showing the loss of sustained positive torques for $q=0.05$ at high $\mathcal{M}$.
  \item In these unequal mass MBHBs ($q = 0.05$), semi-major axis growth is observed across all tested Mach numbers for highly viscous disks (Fig. \ref{fig:adot}), but not for lower viscosity disks (Fig. \ref{fig:adot}). This suggests that CBD interaction alone may be insufficient to shrink such binaries to the GW-driven regime, even in relatively cold disks, unless the viscosity is sufficiently low.
  \item For higher mass ratio binaries ($q \gtrsim 0.1$), the sign of the net binary torque depends on disk temperature: it remains positive in warm (low-$\mathcal{M}$) disks but becomes negative in cold (high-$\mathcal{M}$) disks (Fig. \ref{fig:ell-q}). This transition coincides with changes in both the surface density at the cavity wall and in the accretion dynamics (Figs. \ref{fig:sig-q} and \ref{fig:mdot-1q}). When viscosity is reduced, the switch from net positive to net negative torques occurs at lower Mach numbers for these near equal mass binaries.
  \item In equal mass MBHBs, if the disk has moderate Mach number ($\mathcal{M} \gtrsim 20$), the binaries are efficiently driven toward coalescence within the disk’s lifetime. If instead the viscosity is lower, the binaries still inspiral, largely insensitive to Mach number. 
  \item For near equal mass binaries ($q = 0.25-0.5$) in high viscosity disks ($\nu = \sqrt{2} \times 10^{-3} \ \Omega_b a^2$), accretion onto the secondary black hole dominates over the primary black hole ($f \equiv \dot M_2 / \dot M_1 \ge 1$) when the Mach number is low ($\mathcal{M} \lesssim 20$). At higher Mach numbers, however, the primary black hole begins to accrete more efficiently ($f < 1$). 
  \item In less viscous disks ($\nu = 10^{-4} \ \Omega_b a^2$), the primary black hole dominates accretion for near equal mass binaries ($q = 0.25-0.5$) across all Mach numbers tested due to the secondary black hole rejecting gas from the CBD in inviscid environments. For lower mass ratio binaries ($q \le 0.1$), however, accretion transitions from primary- to secondary-dominated at sufficiently high Mach number. 
\end{enumerate} 

\subsection{Discussion}
\label{subsec:EM}

Our results carry important implications for the MBHB population. Our finding that $q = 0.05$ binaries outspiral in the gas-driven regime, even in high Mach number disks, suggests that low mass ratio systems may not reach the GW-driven phase of evolution unless the disk viscosity is sufficiently low. This raises the possibility that intermediate mass ratio inspirals (IMRIs; $q \simeq 10^{-2}-10^{-4}$) and extreme mass ratio inspirals (EMRIs; $q \lesssim 10^{-4}$) may be rarer than previously expected. Below we discuss how our results inform the potential scarcity of such systems.

For inspiral to occur in unequal mass MBHBs, the secondary, less massive black hole must effectively exert positive torques on the cavity’s inner edge. At mass ratios of $q \simeq 0.05$, the secondary cannot provide sufficient torque, leading instead to outspiral unless the viscosity is as low as $\nu = 10^{-4} \ \Omega_b a^2$. A comparable threshold was identified by \citet{Dittmann2024}, who found that binaries with $0.01 \leq q \leq 0.04$ outspiral up to $\mathcal{M} = 30$. Unlike our results, however, they did not observe a strong viscosity dependence for low-$q$ binaries. Similarly, \citet{Dittmann2022} showed that lower viscosity accelerates inspiral for equal mass binaries. In other recent work, \citet{Derdzinski2021} found that binaries with $10^{-4} < q \leq 10^{-3}$ experience net positive torques (outspiral), while $q = 10^{-4}$ systems show net negative torques up to $\mathcal{M}=30$, though their study did not compute explicit orbital evolution or examine viscosity effects.

Evidence for a similar threshold arises in the planetary migration literature. \citet{Dempsey2021} found that outward migration occurs for super-Jupiters with $q \gtrsim 0.002$ up to $\mathcal{M}=20$. Although their work only probed $q \lesssim 0.02$, these results, together with those from \citet{Derdzinski2021} and \citet{Dittmann2024}, suggest a trend: binaries approaching equal mass can inspiral (in sufficiently cold disks), EMRI-like systems may also inspiral even in low Mach disks, but an intermediate range of mass ratios ($10^{-4} < q \lesssim 0.05$) may be more prone to outspiral if the viscosity is high. Thus, while EMRI formation may remain viable, IMRIs could be significantly suppressed, unless the surrounding disks are sufficiently cold or viscous.

In the high viscosity regime, because the secondary black hole cannot sufficiently repel the CBD at low $q$, gas remains relatively close to the binary even for high Mach numbers (Fig. \ref{fig:sig-q}). The secondary therefore continues to accrete, typically out-accreting the primary black hole, as seen in Fig. \ref{fig:pref-acc}. However, our results in Sec. \ref{subsec:Mdot-results} show that the primary black hole can dominate the accretion especially when $\nu$ is low and $\mathcal{M}$ is high. 

For a binary with $q=0.05$, we find $f \equiv \dot M_2 / \dot M_1 \approx 1.75$. Since the secondary accretes more efficiently, the mass ratio should increase over time, trending toward equal mass as the binary outspirals. If the mass ratio grows to $q=0.1$, our results suggest that the CBD could reverse the torque sign, leading to inspiral regardless of Mach number. Nevertheless, whether such a transition occurs within realistic timescales remains uncertain.

Assuming the binary’s mass-doubling timescale is the Salpeter time, $t_{\rm sal} \equiv M / \dot M \simeq 5 \times 10^7$ yr, the time to double the separation is $t_{2a} = \ln(2) t_{\rm sal} / \epsilon$, where $\epsilon$ is defined via $\dot a_{\rm gas}/a \equiv \epsilon \dot M / M$ \citep{Clyburn2025}. Over one Salpeter time, the separation thus grows by a factor $2^{\epsilon}$. In Fig. \ref{fig:adot}, $\epsilon$ ranges from $\sim 1$–20 for $q=0.05$, depending on the Mach number and we find that lower Mach number disks produce larger $\epsilon$, while higher Mach number disks yield smaller values.

If such a binary evolved in a disk with $\mathcal{M} \simeq 13$, the timescale to reach the GW-driven regime would increase by $2^{20} \simeq 10^6$, drastically extending the system’s lifetime. By contrast, at $\mathcal{M} \simeq 52$, this factor drops to $2^4 = 16$. This implies that low mass ratio MBHBs ($q=0.05$) in cooler disks ($\mathcal{M} \gtrsim 50$) may evolve toward higher mass ratios and eventually inspiral within a reasonable timescale, whereas those in warmer disks ($\mathcal{M} \lesssim 50$) may remain widely separated, potentially never merging. Whether this outcome persists across broader conditions remains an open question and an area of future work.

The above evolutionary picture depends critically on the preferential accretion behavior of low mass ratio systems. To our knowledge, the lowest mass ratios probed to date for preferential accretion are $q = 0.001$ at $\mathcal{M}=10$ \citep{Clyburn2025}, where $f \approx 3$–5 was found for $q = 0.001$–$0.02$. Other studies at higher $q$ (\citealt{Dittmann2024}; \citealt{Farris2014,Duffell2020,Munoz2020,Dittmann2021,Siwek2023a}) report similar trends, with $1 \lesssim f \lesssim 8$. Our results presented in Sec. \ref{subsec:Mdot-results} suggest, however, that the primary black hole can out-accrete the secondary in inviscid disks for near equal mass binaries. Nevertheless, a gap remains in the literature for intermediate mass ratios. Additional simulations will be essential to determine whether such systems maintain preferential accretion and, if so, whether high viscosity disks can still drive them to merge within a Hubble time.

\subsection{Caveats and Future Work}
\label{subsec:future-work}
The simulations presented here are based on two-dimensional, vertically averaged solutions to the Navier–Stokes equations, assuming locally isothermal thermodynamics. The effects of shock heating, radiation, magnetic fields, general relativity, and wind or jet-driven outflows are neglected, and each could influence the results presented here. Our results focus on unequal mass binaries with mass ratios as low as $q = 0.05$; extending this work to lower mass ratios would be necessary to probe the intermediate mass ratio inspiral (IMRI; $q = 10^{-2}–10^{-4}$) regime and to confirm the trend identified here. We also limit our study to Mach numbers of $\mathcal{M} \lesssim 52$, which fall short of the more realistic values expected for thin accretion disks around AGN, $\mathcal{M} = 100 - 1,000$ \cite[e.g.,][]{Shakura1973, Krolik1999, Hubeny2001}. Similarly, our results are constrained by the restricted range of viscosity magnitudes considered; exploring a broader range would be valuable for future work. Despite these simplifications, our simulations provide quantitative estimates of secular inspiral rates and mass ratio evolution, which directly inform the expected population of MBHBs at sub-parsec separations. Such information is crucial for \textit{LISA}, as it provides estimates of the orbital periods and mass ratios of binaries entering the \textit{LISA} band, and thus affects predicted event rates and signal characteristics for the instrument. These results therefore help connect theoretical models of unequal-mass MBHBs in cold, thin disks to the types of sources \textit{LISA} is likely to observe.

%%%%%%%%%%%%%%%%%%%%%%%%%%%%%%%%%%%%%%%%%%%%%%%%%%%%%%%%%%%%%%%%%%%%%%%%%%%%%%%%%%%%%%%%%%%%%%%%%%%%
% Acknowledgements section
%%%%%%%%%%%%%%%%%%%%%%%%%%%%%%%%%%%%%%%%%%%%%%%%%%%%%%%%%%%%%%%%%%%%%%%%%%%%%%%%%%%%%%%%%%%%%%%%%%%%
\section*{Acknowledgements}
M. Clyburn acknowledges the NASA Future Investigators Program (FINESST) support through Award No. 80-NSSC-23K1443. The authors gratefully acknowledge support from the National Science Foundation under Grant Number AST-2408034; and by NASA under award No. 80-NSSC-24K0440. 
All simulations were performed on Clemson University's Palmetto cluster.

%%%%%%%%%%%%%%%%%%%%%%%%%%%%%%%%%%%%%%%%%%%%%%%%%%%%%%%%%%%%%%%%%%%%%%%%%%%%%%%%%%%%%%%%%%%%%%%%%%%%
\section*{Data Availability}
The data underlying this article will be shared on reasonable request to the corresponding author.

%%%%%%%%%%%%%%%%%%%% REFERENCES %%%%%%%%%%%%%%%%%%%%%%%%%%%%%%%%%%%%%%%%%%%%%%%%%%%%%%%%%%%%%%%%%%%%


\bibliographystyle{mnras}
\bibliography{main} % if your bibtex file is called example.bib

%\begin{thebibliography}{99}
%\end{thebibliography}

%%%%%%%%%%%%%%%%%%%%%%%%%%%%%%%%%%%%%%%%%%%%%%%%%%%%%%%%%%%%%%%%%%%%%%%%%%%%%%%%%%%%%%%%%%%%%%%%%%%%

%%%%%%%%%%%%%%%%% APPENDICES %%%%%%%%%%%%%%%%%%%%%%%%%%%%%%%%%%%%%%%%%%%%%%%%%%%%%%%%%%%%%%%%%%%%%%%

\appendix

\section{Sensitivity to Numerical Parameters}
\label{sec:appendix}
We confirm that our results for $\ell$ as a function of Mach number are not significantly affected by numerical parameters such as total number of orbits, sink parameters, domain size, viscosity prescription, and grid resolution.

\begin{figure}
\centering
  \includegraphics[]{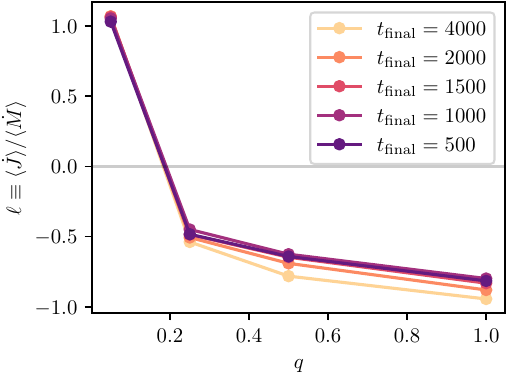}
  \caption{Plot of the dimensionless torque parameter, $\ell$, as a function of mass ratio, $q$, for an MBHB with Mach number $\mathcal{M}=39$. Each line represents the number of orbits averaged over, from a $500$ orbit interval (purple) to a $4,000$ orbit interval (yellow).}
  \label{fig:ell-tfinal}
\end{figure}

In Fig. \ref{fig:ell-tfinal}, we illustrate how $\ell$ depends on the number of binary orbits used in the averaging. Throughout Sec. \ref{sec:results}, we adopt an averaging interval of $1,000$ orbits. For low mass ratio binaries, the torque parameter shows little sensitivity to the choice of averaging interval. In contrast, high mass ratio systems exhibit some dependence; for $q=1.0$, the torque parameter differs by roughly $15\%$ when averaged over $500$ orbits versus $4000$ orbits. This variation may be driven by fluctuations associated with the lump in equal mass binaries. Nevertheless, the results for unequal–mass binaries remain robust.

\begin{figure}
\centering
  \includegraphics[]{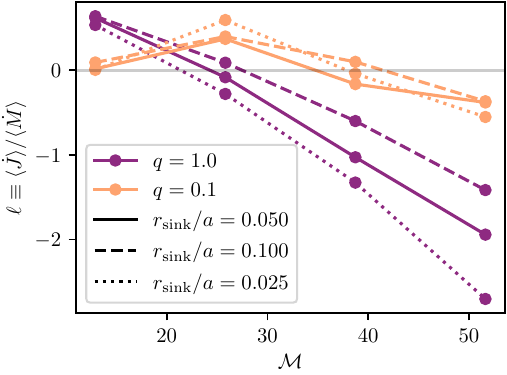}
  \caption{Plot of the dimensionless torque parameter, $\ell$, as a function of Mach number, $\mathcal{M}$, for MBHBs with mass ratio $q=1.0$ (purple) and $q=0.1$ (orange) and varying sink sizes.}
  \label{fig:ell-rs}
\end{figure}

\begin{figure}
\centering
  \includegraphics[]{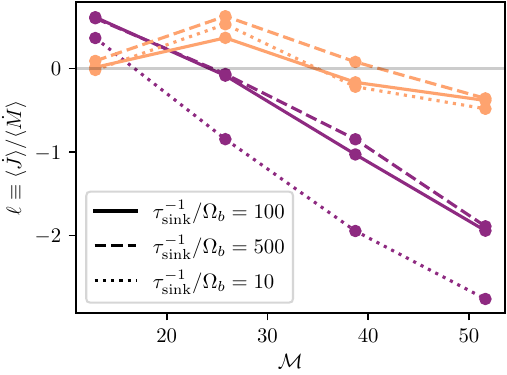}
  \caption{Plot of the dimensionless torque parameter, $\ell$, as a function of Mach number, $\mathcal{M}$, for MBHBs with mass ratio $q=1.0$ (purple) and $q=0.1$ (orange) and varying sink rates. The color for each line corresponds to the mass ratio of the binary, same as in Fig. \ref{fig:ell-rs}.}
  \label{fig:ell-sr}
\end{figure}

Figs. \ref{fig:ell-rs}–\ref{fig:ell-sr} show how $\ell$ depends on sink size and sink rate, respectively, for equal mass ($q=1.0$) and unequal mass ($q=0.1$) binaries. Throughout this work we adopt fiducial sink parameters of $r_{\rm sink} = 0.05 a$ and $\tau_{\rm sink}^{-1} = 100 \ \Omega_b$, consistent with \citet{Tiede2020}. As shown in Fig. \ref{fig:ell-rs}, the torque measurements are largely insensitive to sink size, particularly for the unequal mass binary. For the equal mass case at high Mach numbers, however, the results become more sensitive and fail to converge. Even so, our equal mass results are broadly consistent with previous studies. Since the focus of this manuscript is on unequal mass binaries, where convergence is robust, our main results remain unaffected. 


In Fig. \ref{fig:ell-sr}, we show how the torque measurements depend on the sink rate. We find that the torques are insensitive to sink rate provided it is chosen to be sufficiently high. For all results in this manuscript, we adopt $\tau_{\rm sink}^{-1} = 100 \ \Omega_b$, which ensures convergence of the torque measurements for both equal and unequal mass binaries. At lower sink rates, however, the equal mass case yields artificially low torques, which would incorrectly imply a stronger inspiral at higher Mach number.

\begin{figure}
\centering
  \includegraphics[]{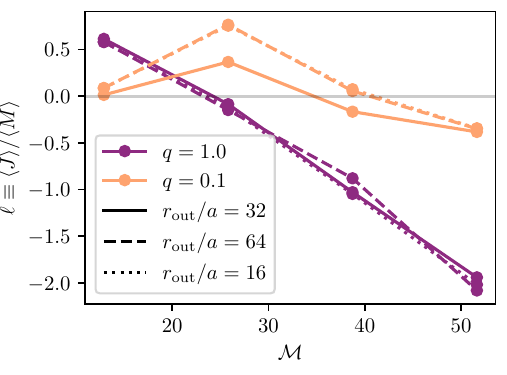}
  \caption{Plot of the dimensionless torque parameter, $\ell$, as a function of Mach number, $\mathcal{M}$, for MBHBs with mass ratio $q=1.0$ (purple) and $q=0.1$ (orange) and varying domain sizes.}
  \label{fig:ell-rout}
\end{figure}

The dependence of $\ell$ on the choice of disk outer radius is illustrated in Fig. \ref{fig:ell-rout}. We find that the measured torques are insensitive to the outer boundary of the computational domain, indicating that the dynamics of interest are well captured within the inner regions of the disk. In Sec. \ref{sec:results} we adopt an outer radius of $r_{\rm out} = 32a$, which we have verified to be sufficiently large to avoid spurious boundary effects and to ensure the accuracy and reliability of our results.

\begin{figure}
\centering
  \includegraphics[]{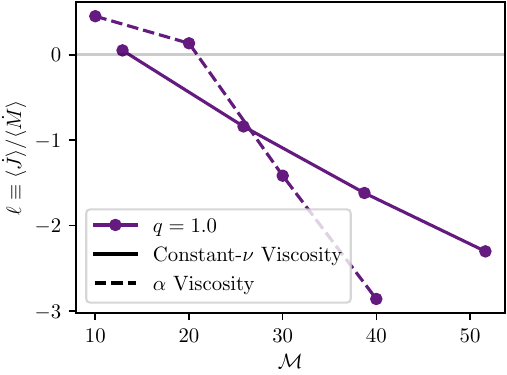}
  \caption{Plot of the dimensionless torque parameter, $\ell$, as a function of Mach number, $\mathcal{M}$, for an MBHB with $q=1.0$ and two viscosity prescriptions shown. The solid lines are for constant-$\nu$ viscosity with $\nu = 10^{-4} \ \Omega_b a^2$ and the dashed lines are for $\alpha$-viscosity with $\alpha = 0.01, 0.04, 0.09, 0.16$ for Mach numbers $\mathcal{M} = 10, 20, 30 ,40$ respectively.}
  \label{fig:ell-visc-alpha}
\end{figure}

Fig. \ref{fig:ell-visc-alpha} illustrates the sensitivity of $\ell$ to the choice of viscosity prescription. Throughout Sec. \ref{sec:results}, we adopt a constant-$\nu$ viscosity for consistency; however, it is also common to model viscosity using an $\alpha$ prescription. To compare the two, we run simulations with a constant-$\nu$ viscosity of $\nu(r) = 10^{-4} \ \Omega_b a^2$ and with $\alpha$-viscosity defined by $\nu(r) = \alpha/\mathcal{M}^2 \sqrt{GMr}$. The parameters $\alpha = 0.01, 0.04, 0.09, 0.16$ are chosen for Mach numbers $\mathcal{M} = 10, 20, 30, 40$, respectively, such that both prescriptions give the same viscosity normalization at $r = a$. We find pronounced differences between the torque measurements: with $\alpha$-viscosity, $\ell$ exhibits a much steeper dependence on Mach number than in the constant-$\nu$ case. This indicates that the strong suppression of accretion at high Mach number reported by \citet{Tiede2025} might not persist under the constant-$\nu$ viscosity prescription.
 
\begin{figure*}
\centering
  \includegraphics[]{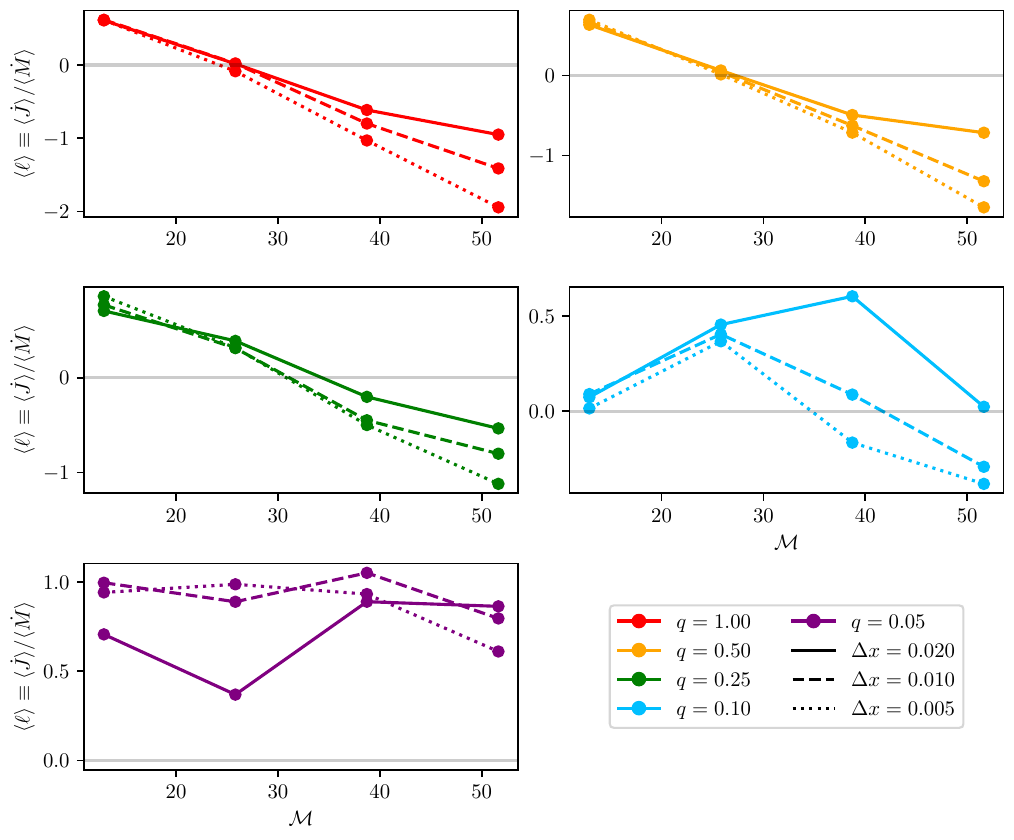}
  \caption{Plot of the dimensionless torque parameter, $\ell$, as a function of Mach number, $\mathcal{M}$, for MBHBs with varying mass ratio and grid spacing. We find convergence of the average torque parameter when the grid spacing is as low as $\Delta x = 0.01 a$ except for some slight deviations at high Mach number.}
  \label{fig:ell-dx}
\end{figure*}

In Fig. \ref{fig:ell-dx} we present the measured torque parameter $\ell$ as a function of Mach number for various binary mass ratios and grid resolutions. For the primary results presented in Sec. \ref{sec:results}, we adopt a  grid spacing with a maximum of $\Delta x = 0.01 a$. This resolution is found to be sufficiently converged, as the differences in torque measurements between $\Delta x = 0.01 a$ and a finer grid of $\Delta x = 0.005 a$ are minimal across all tested Mach numbers and mass ratios. These results indicate that the global torque behavior reported in our simulations is robust to changes in grid resolution.

%%%%%%%%%%%%%%%%%%%%%%%%%%%%%%%%%%%%%%%%%%%%%%%%%%%%%%%%%%%%%%%%%%%%%%%%%%%%%%%%%%%%%%%%%%%%%%%%%%%%

% Don't change these lines
\bsp	% typesetting comment
\label{lastpage}
\end{document}